\documentclass[aps,prl,floatfix,twocolumn]{revtex4}

\usepackage{graphicx,xspace,units}

\newcommand{\micrometer}{\ensuremath{\unit{\mu m}}\xspace}

\begin{document}

\title{Processing Carbon Nanotubes with Holographic Optical Tweezers}

\author{Joseph Plewa}
\author{Evan Tanner}
\author{Daniel M. Mueth}
\affiliation{Arryx, Inc., 316 N. Michigan Ave., Chicago, IL 60601}
\author{David G. Grier}
\affiliation{Dept.~of Physics and Center for Soft Matter Research,
New York University, New York, NY 10003}

\date{\today}

\begin{abstract}
We report the first demonstration that carbon nanotubes can be trapped and manipulated
by optical tweezers.  This observation is surprising because individual nanotubes are
substantially smaller than the wavelength of light, and thus should not be amenable to 
optical trapping.  Even so, nanotube bundles, and perhaps even individual nanotubes,
can be transported at high speeds, deposited onto substrates, untangled, and 
selectively ablated, all with visible
light.  The use of holographic optical tweezers, capable of creating hundreds of independent
traps simultaneously, suggests opportunities for highly parallel nanotube processing with light.
\end{abstract}

\maketitle

Prized for their mechanical, chemical, electrical and optical properties,
carbon nanotubes are archetypal nanotechnological building blocks.
Fully realizing their promise for device and materials applications requires
methods for sorting nanotubes by structure and function, for arranging
them into useful and interesting configurations, and for chemically
processing them once they are in place \cite{dai03}.
This need is particularly pressing if nanotubes are to be integrated
into heterostructures whose other components are incompatible
with available lithographic techniques.
A variety of alternative approaches already have been demonstrated, ranging from single-tube
manipulation with atomic force microscopes and self-assembly guided by
molecular-scale interactions to bulk processing through mixing and
extrusion.
A few of these, such as flow-field alignment, address the need to 
process large numbers of nanotubes over relatively large areas with 
sub-micrometer resolution.
None, however, take advantage of opportunities offered by optical
manipulation, primarily because carbon nanotubes are so much smaller than
the wavelength of light that optical trapping should not be feasible.
Here we demonstrate trapping, deposition, and photochemical
transformation of single-wall carbon nanotubes (SWNTs) using dynamic
holographic optical tweezers (HOTs) \cite{dufresne98,curtis02,grier03}.

Our samples consist of commercial SWNTs (Sigma-Aldrich 519308)
dispersed in water by sonication with 0.5\% 
sodium dodecyl sulfate (SDS) for 30 minutes,
with typical lengths ranging from 100~\unit{nm} to 200~\unit{nm} \cite{oconnell02}.
Although the tubes' characteristic 1.3~\unit{nm} diameters are
two orders of magnitude smaller than the wavelength of visible light,
bundles of tubes still can be detected through conventional dark-field
optical microscopy \cite{yu01}.
Drops of the SWNT dispersion were placed on \#1 glass coverslips and mounted
on the stage of a Nikon TE-2000 microscope integrated into an Arryx BioRyx 200
holographic optical trapping system.
Images were created with a Hamamatsu C7190-23 electron bombardment
charge-coupled device (CCD) camera.

Like conventional optical tweezers \cite{ashkin86}, HOTs
use forces exerted by a strongly focused beam of laser light to trap
mesoscopic objects.
Rather than forming a single trap, however, HOTs use computer-designed
holograms to create arbitrary three-dimensional configurations of
traps from a single input beam \cite{curtis02}.
Updating the hologram in real time updates the
configuration, allowing each trap to move independently
in three dimensions.

\begin{figure}[htbp]
  \centering
  \includegraphics[width=\columnwidth]{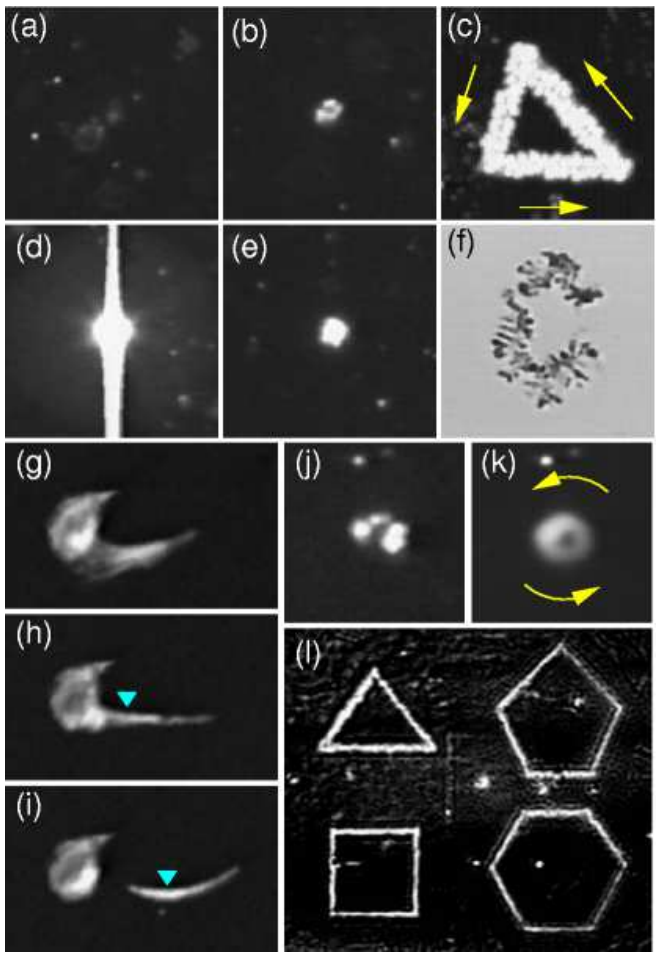}
  \caption{Images of carbon nanotubes processed with holographic optical tweezers.
  (a-c) A dispersion of SWNTs is gathered into an optical trap and translated through water
  at up to 100~\unit{\micrometer/sec}.
  Arrows in (c) indicate the direction of motion.
   A burst of intense illumination that saturates the CCD camera (d) deposits the 
   nanotubes onto a substrate (e).
  (f) Repeating this process forms extended structures.
  (g-i) A single optical tweezer can extract a rope of nanotubes from a bundle.  The arrows in (h)
  and (i) indicate
  the tweezer's position.  (j) and (k) show bundles of SWNTs trapped and spun by an
  optical vortex with helical pitch $\ell = 10$.  (l) Shows four geometric shapes simultaneously
  cut into bucky paper with four holographic optical tweezers.  The square in this figure
  is 2.9~\micrometer across and serves as a scale bar for the images.}
  \label{fig:images}
\end{figure}

Figure~\ref{fig:images}(a) shows a typical snapshot of 
freely diffusing nanotubes, the smallest visible features corresponding
to single nanotubes \cite{yu01}.
Powering a single 30~\unit{mW} optical tweezer at a wavelength of
532~\unit{nm} gathers all the nanotubes in the vicinity to the focus, as in
Fig.~\ref{fig:images}(b).
Our observations suggest that even a single nanotube can be trapped, although
trapping is more robust once large numbers have fallen into the potential well.
To the best of our knowledge, this is the first report of
optical trapping of carbon nanotubes.
Projecting a closely-spaced sequence of traps translates the gathered
bundles through the water, as shown in the time-lapse
image in Fig.~\ref{fig:images}(c).
Turning off the laser trap at this point frees the
particles to diffuse back to their initial random state.

Gathering and moving nanotube bundles is a useful
step toward fabricating structures.
Pressing the trapped bundles onto a substrate such as a
glass surface and briefly increasing the applied laser power
to 50~\unit{mW} (Fig.~\ref{fig:images}(d)) irreversibly
deposits the trapped tubes, as shown in Fig.~\ref{fig:images}(e).
This process can be repeated to create permanent
large-scale structures 
such as the letter ``C'' in Fig.~\ref{fig:images}(f).
This image was created in bright-field illumination.

In addition to manipulating free-floating nanotubes,
optical tweezers also can prise apart
stationary nanotube bundles.  The sequence of images
in Fig.~\ref{fig:images}(g), (h) and (i) shows a nanotube
rope being pulled free from a large bundle with a single optical 
tweezer.  Unlike the other samples, this bundle consisted of
multi-walled nanotubes (MWNTs) (Sigma-Aldrich 406074) dispersed in a 0.5\% SDS
solution without sonication to avoid breaking up the fibers.
The extracted rope can be transported, bent, straightened, and rotated freely in
three dimensions using multiple optical tweezers.

Reshaping the trapping light's wavefronts makes possible
more sophisticated transformations than can be accomplished
with conventional optical tweezers \cite{grier03}.
For example, molding the laser's planar wavefronts
into $\ell$-fold interlocking helices transforms the
optical traps into ring-like optical vortices
capable of exerting torques as well as forces \cite{he95a,simpson96,gahagan96}.
Figure~\ref{fig:images}(j) shows a snapshot of
several nanotube bundles trapped on the periphery
of an $\ell = 10$ optical vortex.  The nanotubes both
absorb and scatter radiation from the beam, thereby
absorbing some of its orbital angular momentum.
The resulting torque causes the bundles to travel
around the ring's circumference, as can be seen in
the time-averaged image in Fig.~\ref{fig:images}(k).

This demonstration of nanotube spinning by photon
orbital angular momentum
complements a recent proposal that photons' spin angular
momentum could be used to twirl nanotubes through Umklapp processes \cite{kral02}.
By contrast, trapping and spinning in optical vortices results
from first-order scattering and absorption, and does not necessarily
disturb the nanotubes' electronic state.
This further suggests that nanotubes are trapped through
conventional optical gradient forces, rather than a more exotic mechanism.

While avoiding nonlinear optical processes can be beneficial,
photochemically induced transformations also are useful for processing
nanotube structures, as demonstrated by the image in Fig.~\ref{fig:images}(l).
Here, a uniform mat of SWNTs, also known as bucky paper, roughly 2~\micrometer
thick was deposited on a glass coverslip and then dried.
The dried mat was illuminated with four optical tweezers, each powered
by 10~\unit{mW} of laser light, which suffices to ablate the nanotubes in air,
but not to ignite them.
In this case, the holographically projected points of light act as
optical scalpels \cite{ashkin87a,berns92a} rather than traps.
Scanning the optical scalpels across the sample created the pattern
of geometric shapes in just over two seconds.
Precise laser sculpting of aligned nanotube
arrays has been demonstrated using single optical tweezers \cite{lim03,cheong03}.
The ability to create multiple holographic scalpels offers opportunities for highly parallel
processing.
Consequently, this approach is complementary to 
dielectrophoretic deposition \cite{nagahara02,suehiro03},
which requires a fluid medium, and offers additional possibilities for post-processing
through gas-phase photochemistry.

While the present communication has focused on manipulation and ablation of
carbon nanotubes, optical processing may hold even further promise.  For instance,
insulating, semiconducting, and metallic SWNTs differ in their optical scattering
characteristics, and so might be amenable to sorting through optical fractionation 
\cite{korda02b,ladavac03}
in large arrays of optical traps.


\end{document}